\documentclass[fleqn,aps,prc,showkeys,12pt,preprint,groupedaddress,showpacs,amsmath,amssymb,floatfix,reprintnumbers,nofootinbib]{revtex4-1}

\usepackage{graphicx}
\usepackage{epsfig}
\usepackage{graphics}
\usepackage{latexsym}
\usepackage{amsmath}
\usepackage{amssymb}
\usepackage{rotating}
\usepackage{bm}
\usepackage{color}
\newcommand{\qmc}[1]{QMC$\pi$-#1}
\newcommand{\s}[1]{$\sigma$ #1}
\newcommand{\om}[1]{$\omega$ #1}
\newcommand{\rh}[1]{$\rho$ #1}
\newcommand{\J}[1]{$\vec{J}$ #1}

\begin{document}

\title[]{ Tensor  and pairing interactions within the QMC energy density functional\color{black}}

\author{K. L. Martinez}
\author{A. W. Thomas}
\affiliation{CSSM and CoEPP, Department of Physics, University of Adelaide, SA 5005 Australia}

\author{P. A. M. Guichon}
\affiliation{IRFU-CEA, Universit\'e Paris-Saclay, F91191 Gif sur Yvette, France}

\author{J. R. Stone}
\affiliation{Department of Physics (Astro), University of Oxford, OX1 3RH United Kingdom}
\affiliation{Department of Physics and Astronomy, University of Tennessee, TN 37996 USA}

\date{\today}

\begin{abstract}

In the latest version of the QMC model, \qmc{III}-T, the density functional is improved to include  the  tensor  component  quadratic in the spin-current and a pairing interaction derived in the QMC framework. Traditional pairing strengths are expressed in terms of the QMC parameters and the parameters of the model optimised. A variety of nuclear observables are calculated with the final set of parameters. The inclusion of the tensor component improves the predictions for ground-state bulk properties, while it has a small effect on the single-particle spectra. Further, its effect on the deformation of selected nuclei is found to improve the energies of doubly-magic nuclei at sphericity. Changes in the energy curves along the Zr chain with increasing deformation are investigated in detail. The new pairing functional is also applied to the study of neutron shell gaps,
 where it  leads to improved predictions for subshell closures in the superheavy region. 

\end{abstract}

\pacs{21.10.-k, 21.10.Dr, 21.10.Ft, 21.60.Jz}
\keywords{nuclear structure, superheavy nuclei, energy density model, quark-meson coupling model }

\maketitle

\section{Introduction}

In this work we study  the effect  of  the tensor component  in the density functional of the Quark-Meson Coupling (QMC) model and we explore the consequences of using  the  pairing interaction  derived from this same model rather than the usual parametrisations. 
The tensor components  of the density functional are not necessarily related to the bare tensor component of the nucleon-nucleon  interaction. The latter is very short ranged and in the QMC  model its effect is ignored  because one assumes that the bags, which describe the quark structure of the nucleon, do not overlap on average. Moreover it is an old  lore in nuclear physics  that the  tensor component of  the rho and pion  exchanges strongly cancel. Because of the small mass of the pion one cannot implement this cancellation in a local density functional. Here the  tensor  components  correspond to the terms of the QMC  density functional which  are quadratic in the spin-current  density (also called spin-tensor) and they arise naturally from the spin dependent part of the QMC effective interaction. They were neglected in previous works  for simplicity and we stress   that they  do not  introduce new parameters, contrary to  other approaches~\cite{Lesinski2007,Zalewski2008,Bender2009,Anguiano2012, Shi2017}.

The QMC model has been successfully applied in nuclear structure studies both for infinite nuclear matter and finite nuclei~\cite{Stone2006, Whittenbury2014, Stone2016, Guichon2018, Martinez2019, Stone2019}. The model self-consistently relates the dynamics of the quark structure of a nucleon to the relativistic mean fields within the nuclear medium. The previous version, \qmc{II}~\cite{Martinez2019}, showed quite satisfactory results in describing even-even nuclei across the nuclear chart, up to the region of superheavies, despite having fewer model parameters. Saturation properties for nuclear matter obtained from \qmc{II} also lay within the acceptable range, with giant-monopole resonances for chosen nuclei also shown to be consistent with available data.

This new  version, \qmc{III}-T, is optimised using the same protocol as in \qmc{II} and with the new parameters we calculate a range of nuclear observables. Most importantly, we investigate the effects of tensor terms on the energies and deformations of selected nuclei as well as the effect on shell gaps of using the QMC-derived nuclear pairing force.

This manuscript is arranged as follows: Section~\ref{theory}  presents  the major developments in the latest \qmc{III}-T EDF; Section~\ref{method} reviews the fitting protocol used to obtain the new set of parameters; Section~\ref{results} presents and discusses the results obtained from the current model; while in Section~\ref{conclude} we present some conclusions and mention some opportunities for future study.

\section{Theoretical framework}\label{theory}
\subsection{The \qmc{III}-T EDF}\label{theory:QMC}
The preceding version, \qmc{II}, was discussed in a recent review \cite{Martinez2019}, while a detailed derivation of the QMC EDF can be found in Ref.~\cite{Guichon2018}. In this section, we focus on the new features incorporated in the current version, \qmc{III}-T, and discuss the corresponding implications for the description of nuclear structure. 

Recall that in \qmc{II} we write the \s field as \s $= \bar{\sigma} + \delta\sigma$, which naturally leads to a classical mean part of the \s field Hamiltonian, $H_{mean}^{\sigma}$ and a fluctuation part $H_{fluc}^{\sigma}$. The effective QMC nucleon mass is expressed as before, as $M_{Q M C} \left(\bar{\sigma} \right) = M - g_\sigma \bar{\sigma} +\frac{d}{2} (g_\sigma\bar{\sigma})^2$, where $g_\sigma$ is the coupling of the nucleon to the \s meson in free space, $d$ is the scalar polarisability and the classical \s field satisfies the wave equation
\begin{equation*}
-\nabla^2\bar{\sigma} + \frac{dV(\bar{\sigma})}{d\bar{\sigma}}=-\langle \frac{\partial K}{\partial \bar{\sigma}}\rangle \, ,
\end{equation*}
where $K$ is the relativistic nucleon kinetic energy, including its mass. The potential $V(\bar{\sigma})$ is expressed as in \qmc{II}, where it adds an additional parameter $\lambda_3$ to account for the self-coupling of the \s meson. One of the main improvements in this new version is that we employ the full expansion for the \s field solution, $g_{\sigma}\bar{\sigma}$, instead of using a Pad\'e approximant. This solution can be explicitly written in terms of the particle density, $\rho$, and the kinetic energy density, $\tau$, as
\begin{equation}
g_{\sigma}\bar{\sigma} = v \left(\rho,\tau,\nabla^2\rho,(\vec{\nabla}\rho)^2 \right)=v_0(\rho) + v_1(\rho)\tau + v_2(\rho)\nabla^2\rho + v_3(\rho)\left(\vec{\nabla}\rho\right)^2 \, ,
\end{equation}
where
\begin{align}
\begin{aligned}
v_0  &= \frac{-(1+G_\sigma d \rho) + \sqrt{(1+G_\sigma d \rho)^2 + 2 G_\sigma^2 \lambda_3 \rho}}{\lambda_3 G_\sigma} \, ,
\\
v_1  &=\frac{-v'_0(\rho)}{2 M^2_{Q M C} \left(v_{0}(\rho) \right)} \, ,
\\
v_2 &= \frac{1}{\lambda_3 G_\sigma v_0(\rho)+(1+d G_\sigma \rho)}\frac{v'_0(\rho)}{m_\sigma^2}{+}\frac{v'_0(\rho)}{4 M^2_{Q M C} \left(v_{0}(\rho) \right)} \, ,
\\
v_3 &= \frac{1}{\lambda_3 G_\sigma v_0(\rho)+(1+d G_\sigma \rho)}\frac{v''_0(\rho)}{m_\sigma^2} \, .
\end{aligned}
\end{align}
As before, the coupling parameter is defined as $G_\sigma= g^2_\sigma/m^2_\sigma$ where the \s meson mass, $m_\sigma$, is taken as a free parameter in the model. Using the expressions for $H_{mean}^{\sigma}$ and $H_{fluc}^{\sigma}$ in Ref.~\cite{Guichon2018} and upon simplification using the new expressions for $g_\sigma \bar{\sigma}$ and $M_{Q M C} \left(\bar{\sigma} \right)$, we then solve for the expectation value of the \s Hamiltonian. 

The new \s contribution to the total QMC Hamiltonian is now expressed as 
\begin{eqnarray*}\label{eq:Hspi3}
\langle H^\sigma_{QMC\pi-III}\rangle
&=& h_0(\rho) + h_4(\rho)\left(J^2_p + J^2_n\right) 
+ \sum_{f=p,n} h_{1} ^ {f} \left(\rho_{p} , \rho_{n} \right)\tau_f \\ \nonumber
&&
+ \sum_{f=p,n} h_{2} ^ {f} \left(\rho_{p} , \rho_{n} \right)\nabla^2\rho_f 
+ \sum_{f,g=p,n} h_{3} ^ {fg} \left(\rho_{p} , \rho_{n} \right)\vec{\nabla}\rho_f\cdot\vec{\nabla}\rho_g \, ,
\end{eqnarray*}
where the coeffients are defined as
\begin{align*}
\begin{aligned}
h_{0} (\rho) & = M_{Q M C} \left(v_{0} \right) \rho + \frac {1} {2 G_{\sigma}} v_{0} ^ {2} + \frac{\lambda_{3}}{{3!}} v_{0} ^ {3} {+ \frac {1} {4} G_\sigma (1-dv_0)^2 \left(\rho_{p} ^ {2} + \rho_{n} ^ {2} \right)} \, ,
\\
h_{1} ^ {f} \left(\rho_{p} , \rho_{n} \right) & = \frac {1} {2 M_{Q M C} \left(v_{0} \right)} -\frac {1} {4}  \left[\frac{2dv_1G_\sigma (1-dv_0)^2}{1-dv_0}\right] \left(\rho_{p} ^ {2} + \rho_{n} ^ {2} \right) - \frac {1} {2} q (\rho) \rho_{f} \, ,
\\
h_{2} ^ {f} \left(\rho_{p} , \rho_{n} \right) & = - \frac {1} {4 M_{Q M C} \left(v_{0} \right)} - \frac {1} {4}\left[\frac{2dv_2G_\sigma (1-dv_0)^2}{1-dv_0}\right] \left(\rho_{p} ^ {2} + \rho_{n} ^ {2} \right) + \frac {1} {4} q (\rho) \rho_{f} \, ,
\\
h_{3} ^ {f g} \left(\rho_{p} , \rho_{n} \right) & = \frac {v_{0} ^ {'2}} {2 m_{\sigma} ^ {2} G_{\sigma}} - \frac {1} {4} \left[\frac{2dv_3G_\sigma (1-dv_0)^2}{1-dv_0} + {p ^ {\prime 2}} \right] \left(\rho_{p} ^ {2} + \rho_{n} ^ {2} \right) + \delta (f , g) \frac {1} {8} p ^ {2}  \, ,
\\ 
h_{4} (\rho) & = \frac {1} {4} p ^ {2} \, ,
\end{aligned}
\end{align*}
with $p (\rho)  = \frac {{-\sqrt{G_{\sigma}}} \left(1 - d v_{0} \right)} {{m_\sigma}}$ and $
q (\rho) = \left(1 + \frac {m_{\sigma} ^ {2} } {2 M{^2}_{Q M C} \left(v_{0} \right)} \right) p ^ {2}$. 

Additional contributions to the spin-independent part of the QMC Hamiltonian come from the  \om and \rh vector mesons, where we define the coupling parameters $G_\omega=g^2_\omega/m_\omega^2$ and $G_\rho=g^2_\rho/m_\rho^2$, with the masses taken at their physical values. There are also spin-dependent contributions to the spin-orbit (SO) terms of the Hamiltonian and the central \om and \rh as well as the SO parts are treated in the same way as in \qmc{II}.

The total QMC Hamiltonian is solved in a Slater determinant by filling the single-particle states $\{\phi\}$ up to a Fermi level corresponding to the number of protons, $Z$, and neutrons, $N$, in a given nucleus. The densities are defined as before as:
\begin{eqnarray}
\rho_{m}(\vec{r}) & = & \sum_{i\in F_{m}}\sum_{\sigma}\left|\phi^{i}(\vec{r},\sigma,m)\right|^{2},\,\,\, \nonumber \rho=\rho_{p}+\rho_{n} \, ,\\ \nonumber
\tau_{m}(\vec{r}) & = & \sum_{i\in F_{m}}\sum_{\sigma}\left|\vec{\nabla}\phi^{i*}(\vec{r},\sigma,m)\right|^{2},\,\,\,\tau=\tau_{p}+\tau_{n} \, ,\\  \nonumber
\vec{J}_{m} & = & i\,\sum_{i\in F_{m}}\sum_{\sigma\sigma'}\vec{\sigma}_{\sigma'\sigma}\times\left[\vec{\nabla}\phi^{i}(\vec{r},\sigma,m)\right]\phi^{i*}(\vec{r},\sigma',m),\,\,\,\vec{J}=\vec{J}_{p}+\vec{J}_{n} \, , 
\end{eqnarray}
where $\rho$, $\tau$, and \J are the particle, kinetic and spin-tensor densities, respectively. In \qmc{III}-T, we take all tensor terms (i.e. quadratic in \J) appearing in the total QMC functional. These additional terms are discussed in the next subsection.
Finally, we note that the spin-orbit piece of the Hamiltonian, $H_{SO}$, is identical to that used in \qmc{II}. It includes both the time and space components of the meson-nucleon couplings.

\subsection{Tensor contribution within the QMC model}\label{theory:tensor}
In traditional mean-field calculations the tensor terms are often neglected. This was the case in the previous versions of the QMC model where the quadratic \J terms were set to zero. The effect of these terms may be small but since they naturally arise in the QMC model and are fully expressed in terms of the existing parameters, without any serious complication in the functional, we include them in \qmc{III}-T. The tensor terms arising from the time component of the meson fields can be written as
\begin{equation}
\label{eq:centralJ}
H_{\sigma,\omega,\rho}^J = \left(\frac{G_\sigma (1-dv_0)^2}{4 m_\sigma^2} - \frac{G_\omega}{4 m_\omega^2}\right) \sum_m \vec{J}^2_m - \frac{G_\rho}{4 m_\rho^2} \sum_{m,m'} S_{m,m'} \vec{J}_m \cdot \vec{J}_{m'} \, , 
\end{equation}
where $S_{m,m'}=\delta_{m,m'} m^2 + \frac12 (\delta_{m,m'+1} + \delta_{m',m+1})$. For the like-particle tensor component, we can see a strong cancellation between the \s and \om contributions which is further decreased by the \rh term.

The additional tensor terms arising from the relativistic spin-dependent part of the model are expressed as
\begin{equation}
\label{eq:SOJ}
H_{S}^J = -\frac{G_\sigma-G_\omega}{16 M^2}\sum_m \vec{J}^2_m + \frac{G_\rho}{16M^2} \sum_{mm'} S_{m,m'} \vec{J}_m \cdot \vec{J}_{m'} \, . 
\end{equation}
Again, we see a strong cancellation in the like-particle component between the \s and \om contributions, while \rh appears with an opposite sign. Because of these cancellations the tensor terms in \qmc{III}-T are expected to make a relatively small overall contribution to the total QMC Hamiltonian.

\subsection{The pairing functional}
\label{sub:QMCpair}
In the standard treatment for pairing energy, it is common to take either a $\delta$-function  force that is constant throughout the nuclear volume (DF) or a density-dependent $\delta$ interaction (DDDI) which is concentrated on the nuclear surface, or both (mixed pairing). The pairing potential can be expressed as
\begin{equation}
\label{pair}
V_{pair} = - V_{p,n} \left[1- \left(\frac{\rho}{\rho_c}\right)^\alpha \right] 
\delta(\vec{r} - \vec{r}') \, ,
\end{equation}
where $V_{p,n}$ are the proton and neutron pairing strength parameters. For DF pairing, the critical density, $\rho_c$, is set to $\infty$, while it is usually chosen to be equal to the saturation density, $\rho_0 = 0.16$ fm$^{-3}$, for DDDI. In some other cases, $\rho_c$ is taken to be a free parameter. The power $\alpha$ is an additional parameter which controls the density-dependence for mixed pairing. For DDDI, $\alpha$ is simply set to 1.0. At the most, one has to fit four extra parameters: $V_p$, $V_n$, $\rho_c$ and $\alpha$, for the pairing functional in addition to the parameters of the mean-field Hamiltonian.

Within the QMC framework, the pairing force can be seen as the interaction between nucleons modified by medium effects. In the same way as the HF potential is treated in the Bogoliubov theory, we can compute the pairing potential with the QMC Hamiltonian as
\begin{equation}
\label{QMCpair}
V_{pair}^{QMC} = -\left(\frac{G_\sigma}{1+ d' G_\sigma \rho(\vec{r})} - G_\omega - \frac{G_\rho}{4} \right) \delta (\vec{r}-\vec{r}') \, , 
\end{equation}
where we have the modification, $d' = d + \frac13 G_\sigma \lambda_3$, as the result of the cubic self-interaction of the \s meson. With this expression for the pairing interaction, we now do away with the additional pairing parameters which appear in Eq.~(\ref{pair}). The QMC-derived pairing potential in Eq.~(\ref{QMCpair}) is fully expressed in terms of the existing parameters of the model,  which are fitted together with the mean-field part of the QMC Hamiltonian.

Other contributions to the total QMC EDF are the single-pion exchange, which is evaluated using local density approximation and the Coulomb interaction, which is expressed in a standard form including its direct and exchange terms. These functionals are taken as in \qmc{II} and the reader is referred to Ref.~\cite{Guichon2018} for more discussion.

\section{Method}
\label{method}
The QMC Hamiltonian for finite nuclei is solved using an HF+BCS code SkyAx which allows for axially-symmetric and reflection-asymmetric shapes ~\cite{skyax}. Once the densities are computed, nuclear observables such as binding energies $BE$ and \textit{rms} charge radii $R_{ch}$ can be   obtained for a given nucleus. 

To optimise the \qmc{III}-T functional, a derivative-free optimisation algorithm known as POUNDeRS~\cite{petsc-user-ref,petsc-efficient,tao-user-ref} has been employed. There are a total of five parameters to fit to data, consisting of the three couplings, $G_\sigma$, $G_\omega$, and $G_\rho$, the \s self-coupling parameter, $\lambda_3$, and the \s meson mass, $m_\sigma$. The same set of seventy magic nuclei, just as in \qmc{II} optimisation, were included in the fit. For the present fit, however, we only include available data for $BE$ and $R_{ch}$ giving a total of 129 data points. The objective function to be minimised is defined as
\begin{equation*}
F(\mathbf{\hat{x}}) = \sum_i^{n}\sum_j^{o} \left(\frac{\bar{s}_{ij}-s_{ij}}{w_j}\right)^2 \, ,
\label{objfxn}
\end{equation*}
where $n$ is the total number of nuclei, $o$ is the total number of observables and $s_{ij}$ and $\bar{s}_{ij}$ are the experimental and fitted values, respectively. $w_j$ stands for the \textit{effective} error for each observable, set in this fit to be  1 MeV for $BE$ and 0.02 fm for $R_{ch}$ for all nuclei. We use the \qmc{II} parameter set from Ref.~\cite{Martinez2019} as the starting point of the parameter search. The corresponding nuclear matter properties (NMP) were expected to be in the same range as in \qmc{II} and that indeed is the case. With the final parameter set for \qmc{III}-T, we calculate various nuclear observables which are discussed in the next section.

\section{Results and discussion}
\label{results}
In this section, we present and discuss the results from \qmc{III}-T EDF mainly in view of: 1) the effect of adding the tensor component to the functional and 2) using the QMC-derived pairing functional.

\subsection{Effects of tensor terms}
In this subsection, we investigate the effects of tensor component within \qmc{III}-T. Table~\ref{pars} shows the parameters for the cases with tensor contribution (labelled `\qmc{III}-T') and without tensor (labelled `\qmc{III}'), along with their corresponding NMPs. Notice that the final parameters did not change much with the addition of tensor terms; basically the coupling parameters are slightly reduced while both $m_\sigma$ and $\lambda_3$ remain unchanged.  The resulting NMPs are also almost the same for both cases, with and without the tensor term. The effects on masses and single-particle energies for finite nuclei, however, can be quite different as will be presented in the succeeding results.

\begin{table}[th!]
\begin{ruledtabular}
	\caption{Parameters of \qmc{III} with and without tensor component and corresponding NMPs along with their errors (written in parentheses).}
	\centering
	\begin{tabular} { c c c c c | c c c c c}
	Parameter&\multicolumn{2}{l}{\qmc{III}-T}&\multicolumn{2}{l}{\qmc{III}}&NMP&\multicolumn{2}{l}{\qmc{III}-T}&\multicolumn{2}{l}{\qmc{III}}\\\hline
	$G_\sigma$ [fm$^{-2}$]&9.62 &(0.01) & 9.66 & (0.02)&$\rho_0$ [fm$^{-3}$]&0.15&(0.01)& 0.15&(0.01) \\
	$G_\omega$ [fm$^{-2}$]&5.21&(0.01)& 5.28 & (0.01)&$E_0$ [MeV]&-15.7&(0.2)&-15.7&(0.2)\\
	$G_\rho$ [fm$^{-2}$]	&4.71&(0.03)& 4.75 & (0.03)&$a_{sym}$ [MeV]&29&(1)&29&(1)\\
	$m_\sigma$ [MeV]		&504 &(1)& 504 & (1)&	$L_0$ [MeV]	&43&(4)&43&(7)\\
	$\lambda_3$ [fm$^{-1}$]&0.05&(0.01)&0.05 & (0.01)&$K_0$ [MeV]	&233&(2)&235&(2)\\
	\end{tabular}
	\label{pars}
\end{ruledtabular}
\end{table}

\begin{center}
\begin{figure}[th!]
\includegraphics[width=0.35\textwidth]{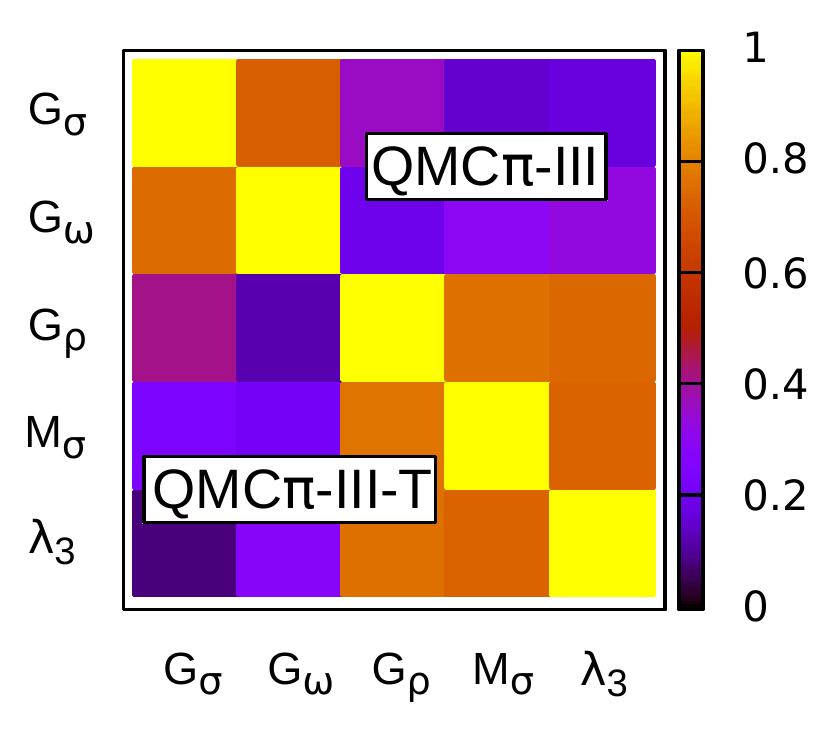}
\caption{Comparison between correlation matrices of the \qmc{III}-T and \qmc{III} parameters. Black represents no correlation while yellow means the parameters are 100\% correlated.}
\label{fig:covmat}
\end{figure}
\end{center}

Figure~\ref{fig:covmat} shows the correlation matrices for both \qmc{III}-T and \qmc{III}. It can be seen that the correlation between any two parameters is very similar for both cases. There is a relatively higher correlation between $G_\sigma$ and $G_\omega$ but both parameters have only a small correlation with the other parameters. Meanwhile, $G_\rho$ is highly correlated with both $m_\sigma$ and $\lambda_3$ and, just as in \qmc{II}, the \s meson mass also has high correlation with $\lambda_3$.
\subsubsection{Masses and radii across the nuclear chart}
Using the final parameter sets presented in Table \ref{pars}, we calculate the energies and radii of known even-even nuclei across the nuclear chart. The same was done in the previous QMC versions and their results are added here for comparison.

\begin{figure}[tbh!]
\includegraphics[width=1.0\textwidth]{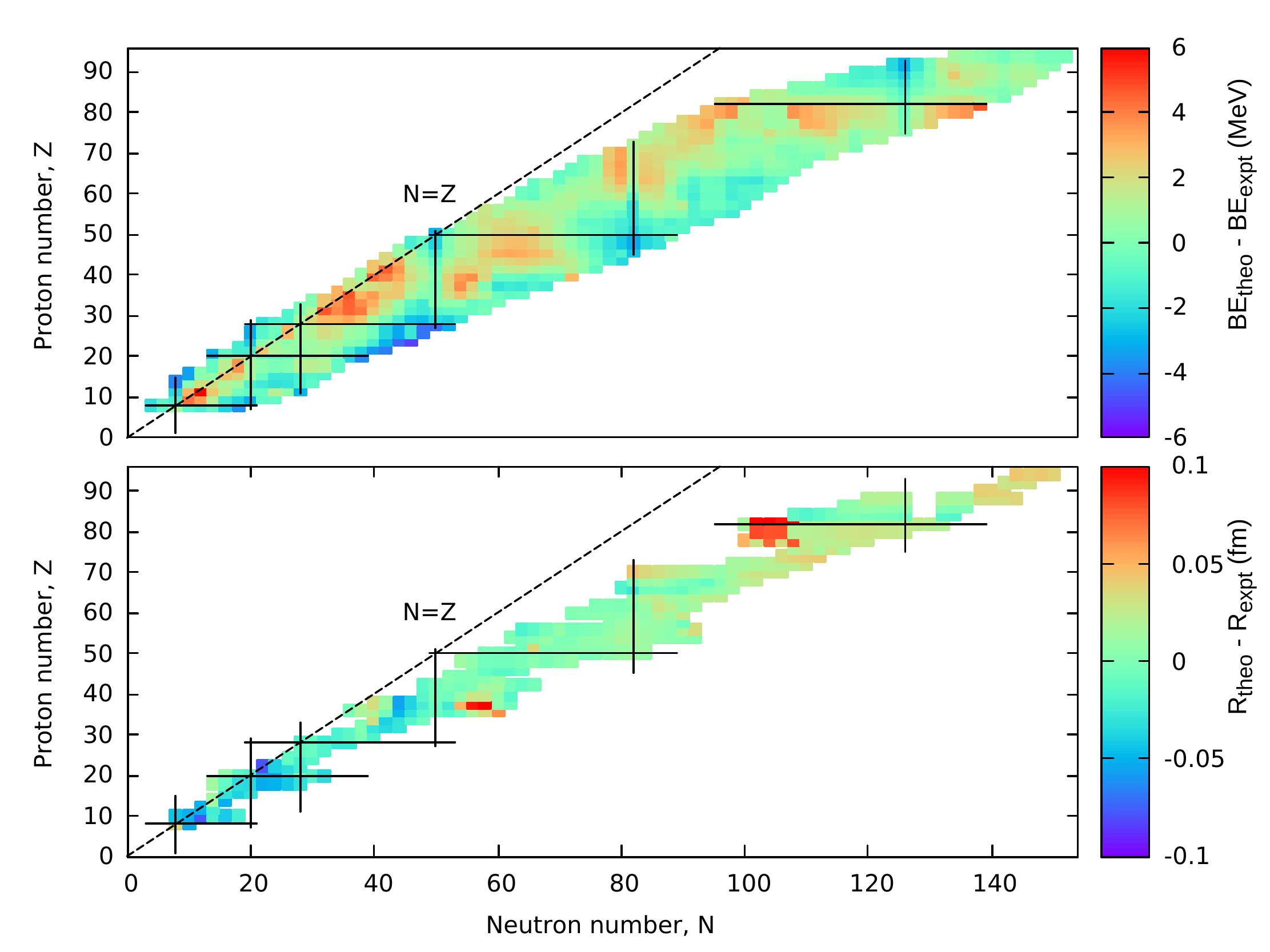}
\caption{$BE$ and $R_{ch}$ residuals for known even-even nuclei with $Z < 96$ computed from the \qmc{III}-T functional. Atomic mass data used to compute the binding energy residuals are taken from Ref.~\cite{Wang2017} and $rms$ charge radii data are from Ref.~\cite{Angeli2013}. Nuclei with magic numbers are indicated by solid lines and symmetric nuclei ($Z=N$) are shown in a dashed line.}
\label{fig:resid}
\end{figure}

Figure \ref{fig:resid} shows the residuals for $BE$ and $R_{ch}$ obtained from \qmc{III}-T. As 
in \qmc{II}~\cite{Martinez2019}, there are relatively large residuals along the symmetric line $N=Z$. This may be attributed to the Wigner energy, the contribution of which is conventionally discarded in mean-field theories. The $BE$ residuals in \qmc{III}-T vary as much as $\pm 6$ MeV, whereas the variation was as large as 8 MeV in \qmc{II}. The $R_{ch}$ residuals, however, remain in the same range at around $\pm 0.1$ fm.

Table~\ref{tab:resid} shows a comparison of rms residuals from various QMC versions, along with results from other nuclear models, for the nuclei included in Figure~\ref{fig:resid}. There are a total of 746 nuclei with known $BE$ and 346 nuclei with known $R_{ch}$ included in the plot. The predictions for $BE$ are greatly improved in \qmc{III}-T compared to the results of \qmc{II}, especially with the addition of the tensor terms. Predictions for $R_{ch}$, on the other hand, remain almost the same and are not much affected with the inclusion of the tensor component. Overall, QMC predictions are comparable to those of the other models, even with a significantly smaller number of model parameters.
\begin{table}[th!]
\begin{ruledtabular}
	\caption{Comparison of $BE$ and $R_{ch}$ residuals from \qmc{III}-T, \qmc{III}, \qmc{II}~\cite{Martinez2019}, Skyrme forces SV-min~\cite{Klup2009} and UNEDF1~\cite{Kort2012}, and FRDM~\cite{Moller2016}.}
	\centering
	\begin{tabular} { c c c c c c c}
	Observable		&\qmc{III}-T&\qmc{III}&\qmc{II}&SV-min&UNEDF1&FRDM\\\hline
        \textit{BE} (MeV)	& 1.74 &2.17 &2.39 & 3.11 &2.14 &0.69\\
        \textit{R$_{ch}$} (fm$^{-3}$)&0.028 &0.028 &0.026& 0.023 &0.027& not available
	\end{tabular}
\label{tab:resid}
\end{ruledtabular}
\end{table}

\subsubsection{Masses along isotopic and isotonic chains}
\begin{figure}[th!] 
\includegraphics[width=0.9\textwidth]{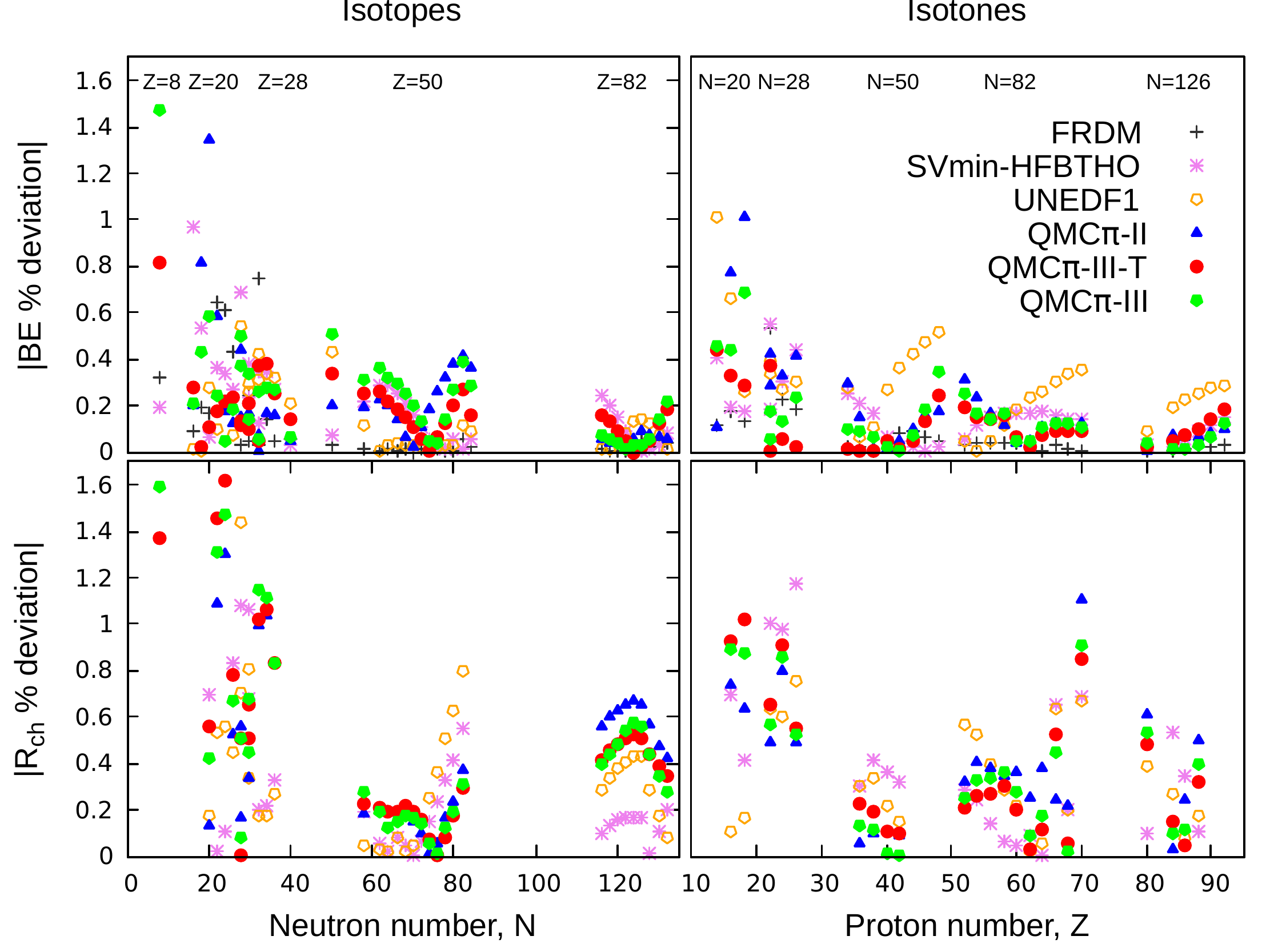}
\caption{Percentage deviation from experiment for binding energies $BE$ and \textit{rms} charge radii $R_{ch}$ for semi-magic isotopes and isotones included in the \qmc{III} fitting procedure. Added for comparison are results for the Skyrme type forces SV-min~\cite{Klup2009} and UNEDF1~\cite{Kort2012}, the finite-range droplet model (FRDM)~\cite{Moller2016} and results from the previous \qmc{II} version~\cite{Martinez2019}. The plot legend is located in the top right panel.}
\label{fig:devs_Jcomp}
\end{figure}

We now look more closely at the effects of adding tensor terms in the \qmc{III}-T functional by comparing the results for the energies and radii of magic isotopes and isotones. Figure~\ref{fig:devs_Jcomp}
shows the fit results obtained from QMC along with results for the same set of nuclei from other nuclear models. Overall, the deviations for these nuclei within the QMC model are in the same range as other nuclear models, particularly having relatively higher values in light to medium nuclei. Within the QMC model, we can see improvements with the \qmc{III}-T version in both energies and radii, with the exception of some energies in the lead chain and radii in light isotones, where \qmc{II} seems to perform better. The tensor effect within \qmc{III}-T is further investigated in the succeeding plot.

\begin{figure}[th!] 
\includegraphics[width=1.0\textwidth]{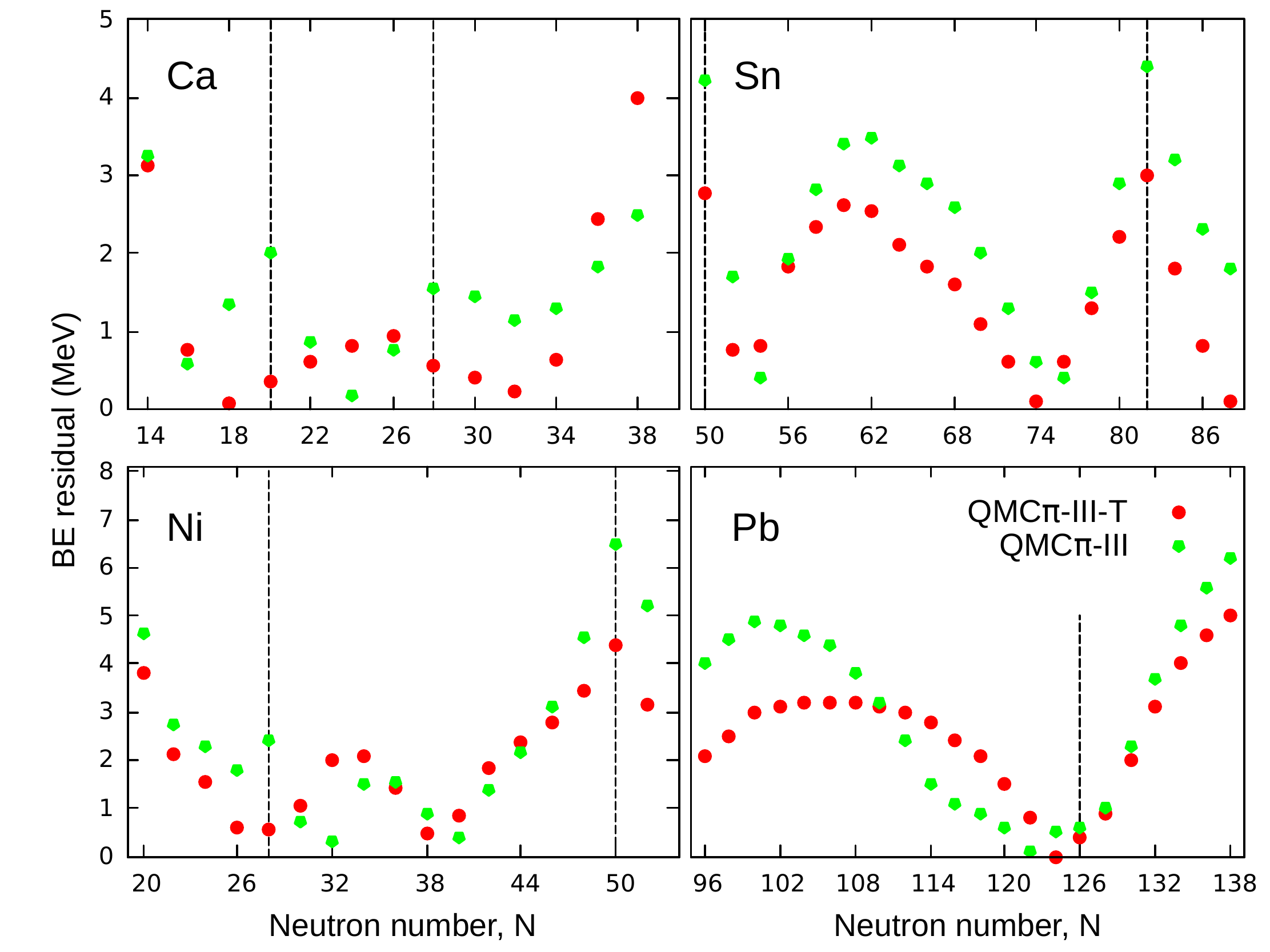}
\caption{Comparison of $BE$ residuals along Ca, Ni, Sn, and Pb chains for the cases with and without tensor contribution. The magic numbers are indicated by dashed lines in the figure where nuclei are doubly-magic. The plot legend is located at the bottom right panel.}
\label{fig:BEresid}
\end{figure}

Considering the results from \qmc{III} with and without tensor component, Figure~\ref{fig:BEresid} compares the $BE$ residuals along the isotopic chains of calcium, nickel, tin and lead. It can be seen that, in general, the inclusion of tensor terms improved the energies for these chains. Only for neutron-rich $^{56,58}$Ca, around $^{60}$Ni, and from $^{194}$Pb towards $^{204}$Pb are the residuals better for the case where the tensor component is neglected. We emphasise that for doubly-magic nuclei, $^{40,48}$Ca, $^{56,78}$Ni and $^{100,132}$Sn, shown with dashed lines in the figure, the tensor component improved the values for total binding energies. For $^{208}$Pb, the effect of the tensor component for $BE$ is not significant. In the next subsection, we tackle the single-particle states of doubly-magic nuclei and how the levels are affected by the tensor component.

\subsubsection{Single-particle states}
\begin{figure}[th!] 
\includegraphics[width=0.75\textwidth]{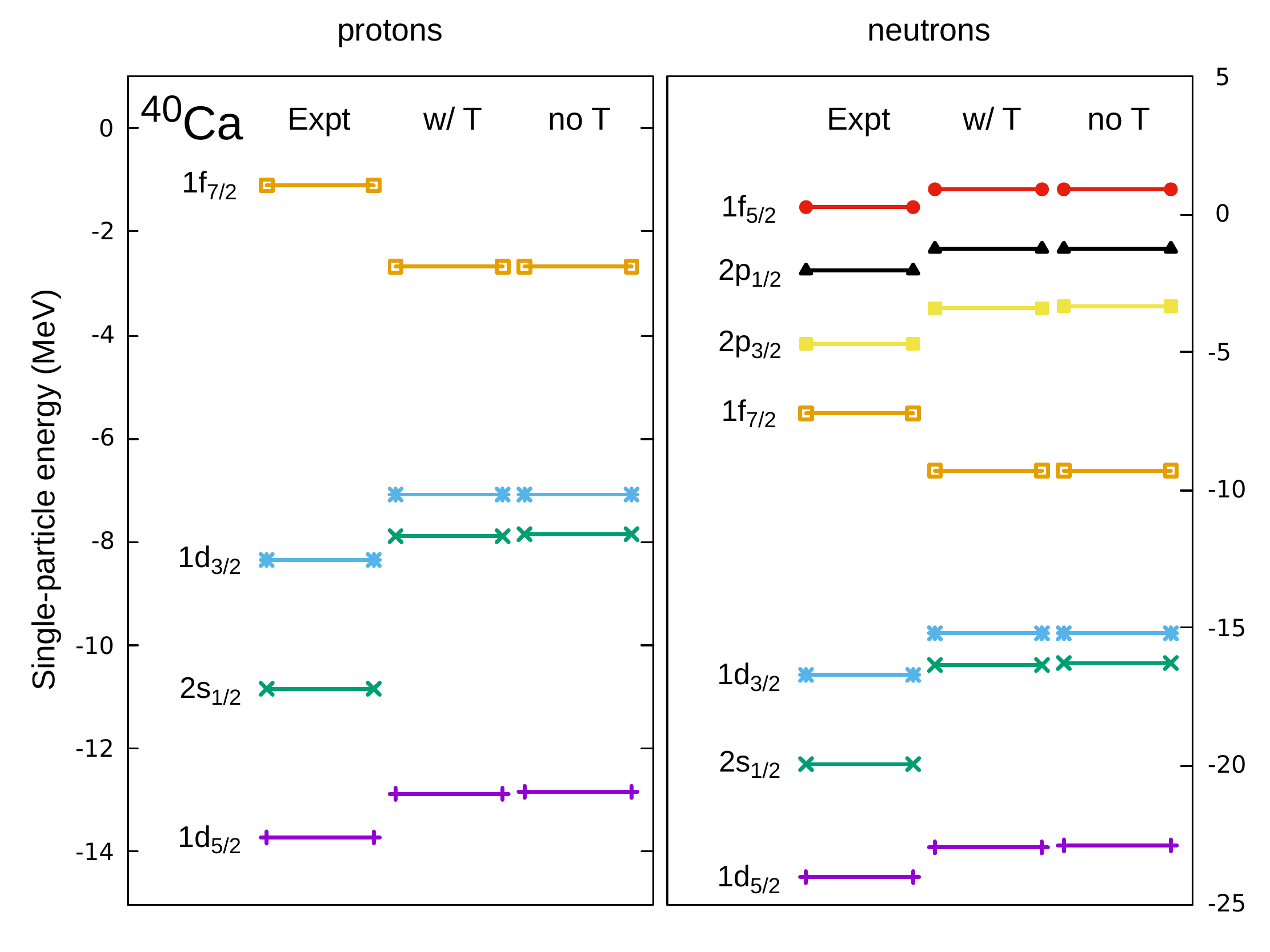}
\caption{Proton and neutron single-particle states for $^{40}$Ca obtained from \qmc{III} with and without the tensor component. Experimental data is taken from~\cite{Grawe2007}. Single-particle levels are shown in different colors and labels are placed before the experimental data for each level.}
\label{Ca40_sps}
\end{figure}

\begin{figure}[th!] 
\includegraphics[width=0.75\textwidth]{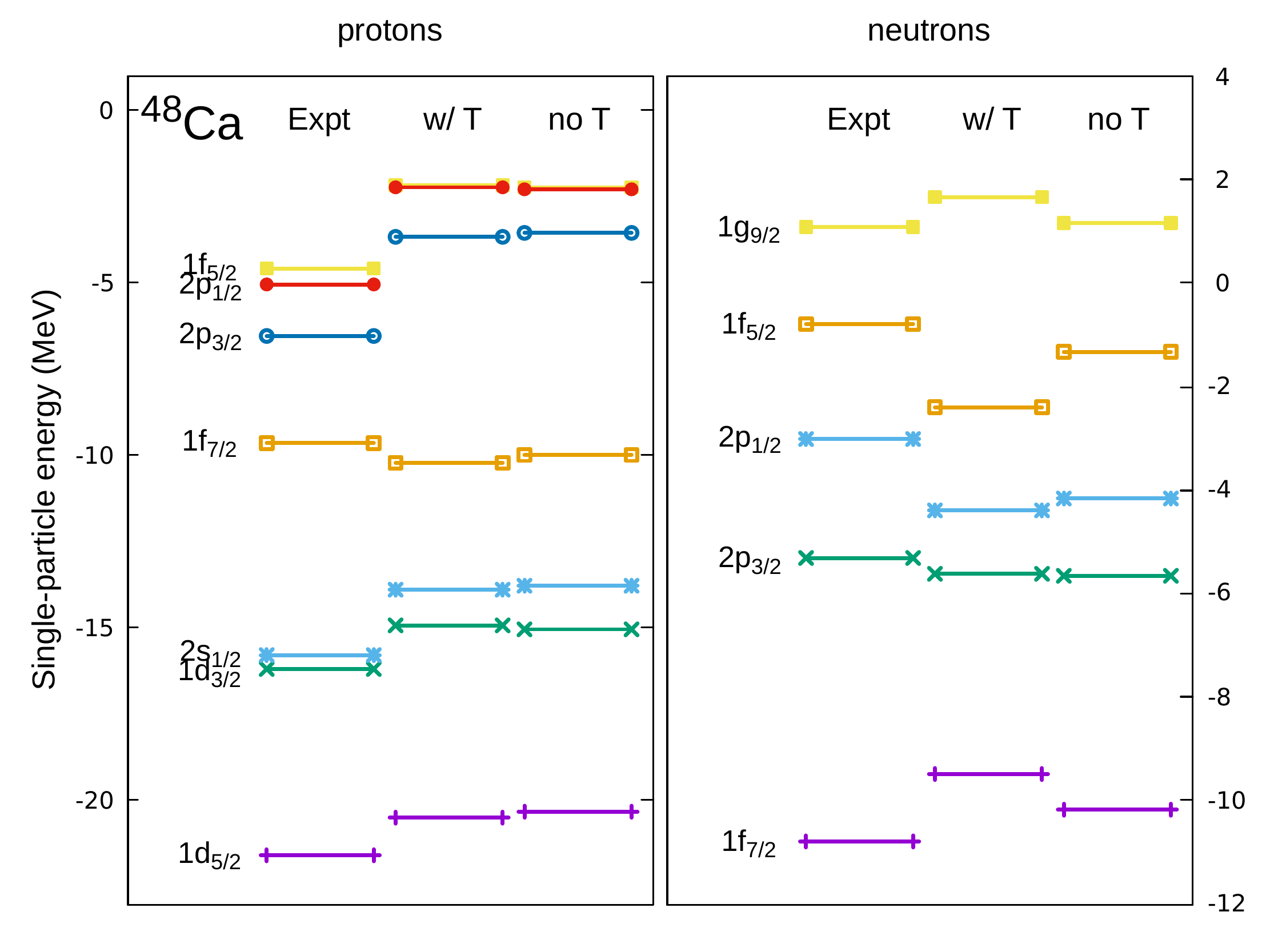}
\caption{Same as in Figure~\ref{Ca40_sps} but for $^{48}$Ca.}
\label{Ca48_sps}
\end{figure}

\begin{figure}[th!] 
\includegraphics[width=0.75\textwidth]{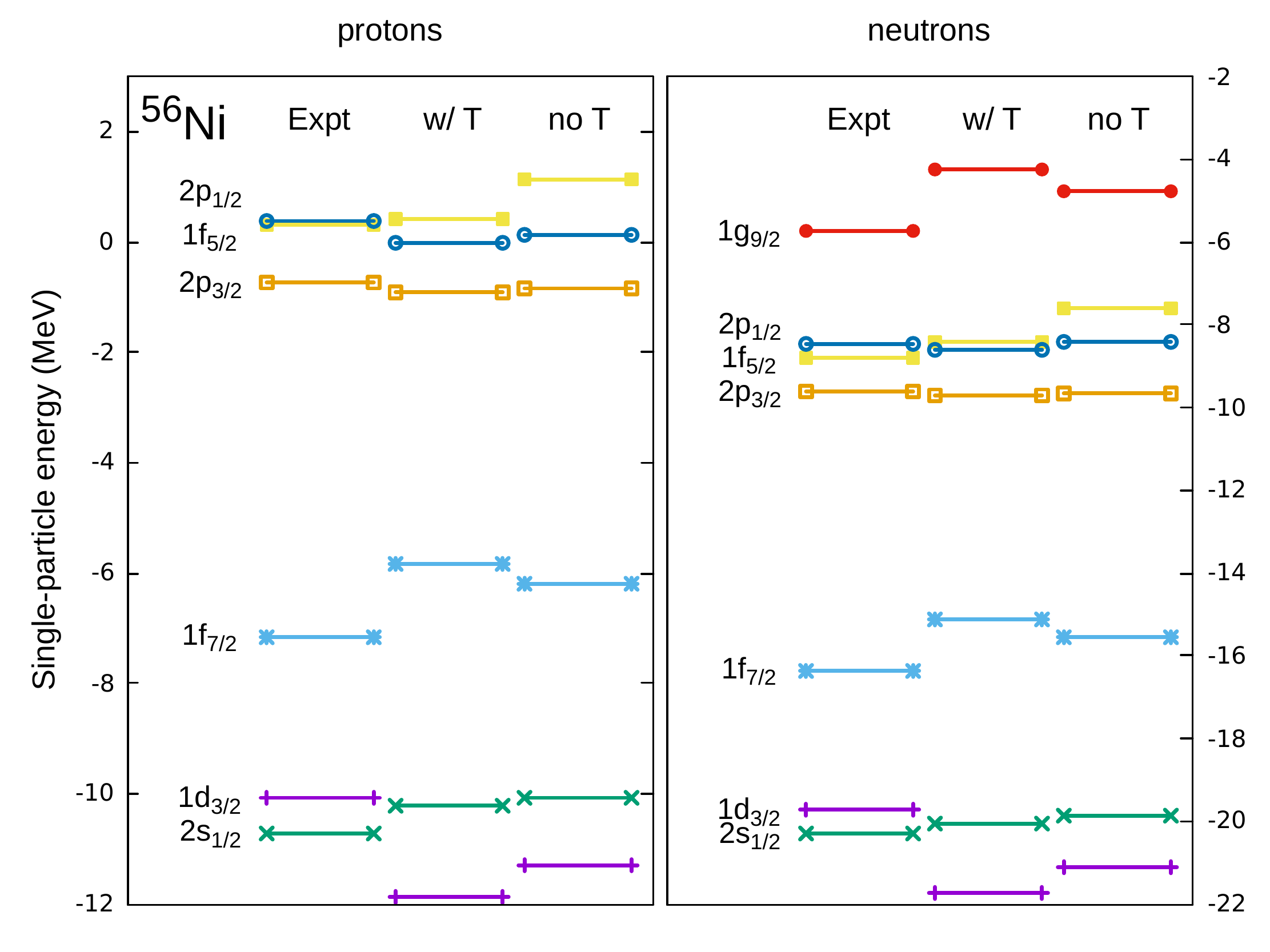}
\caption{Same as in Figure~\ref{Ca40_sps} but for $^{56}$Ni.}
\label{Ni56_sps}
\end{figure}

\begin{figure}[th!] 
\includegraphics[width=0.75\textwidth]{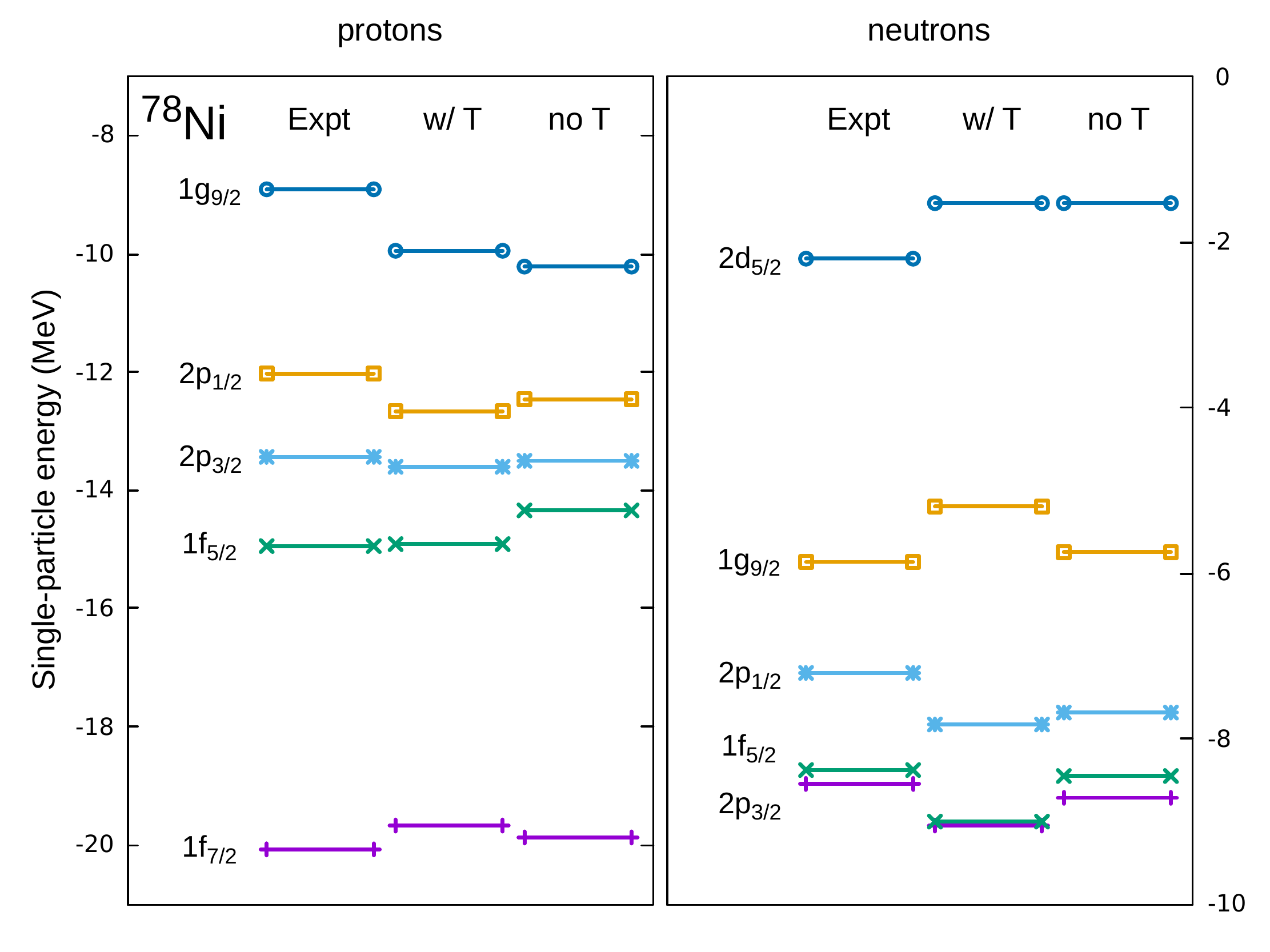}
\caption{Same as in Figure~\ref{Ca40_sps} but for $^{78}$Ni.}
\label{Ni78_sps}
\end{figure}

\begin{figure}[th!] 
\includegraphics[width=0.75\textwidth]{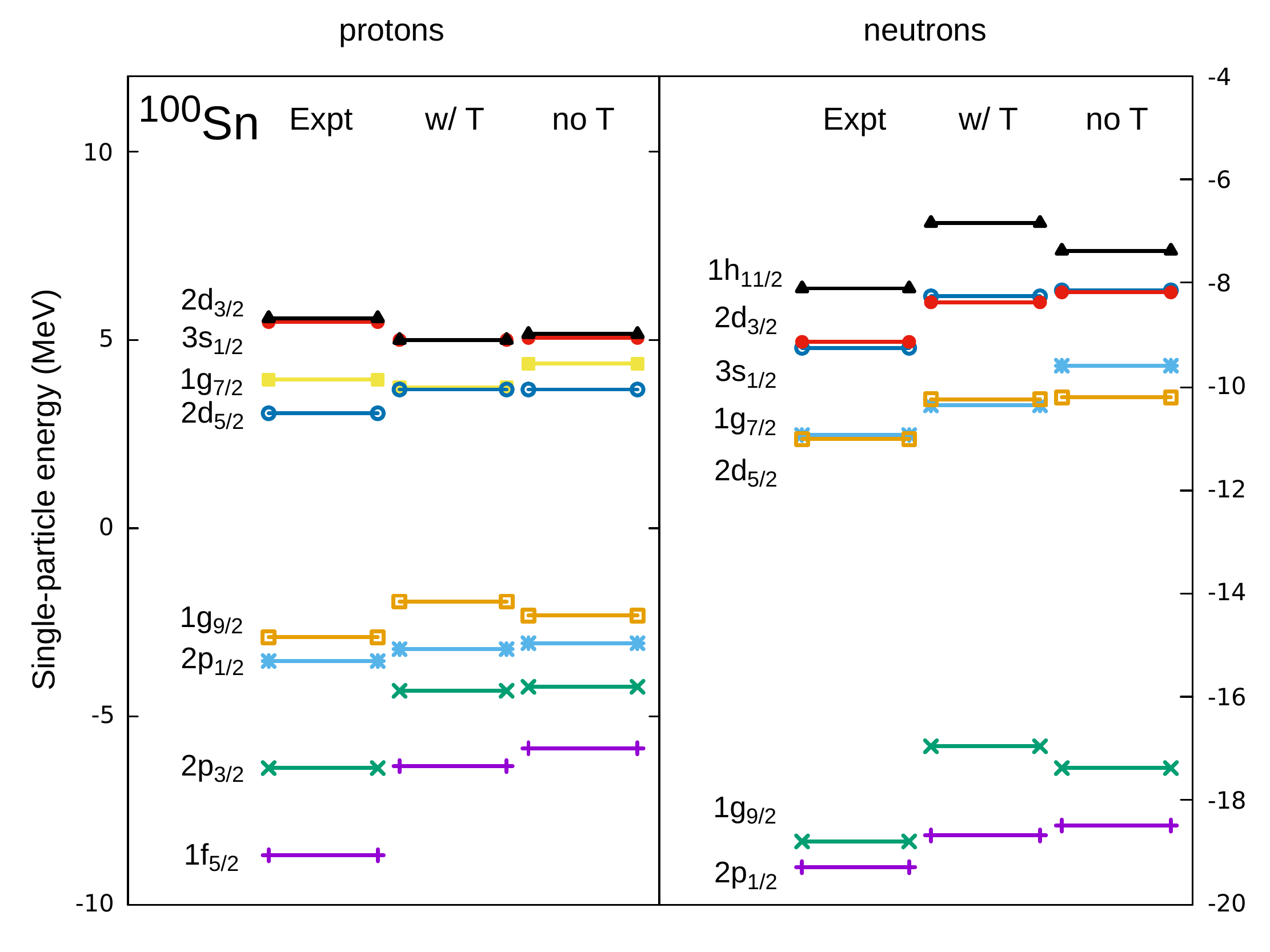}
\caption{Same as in Figure~\ref{Ca40_sps} but for $^{100}$Sn.}
\label{Sn100_sps}
\end{figure}

\begin{figure}[th!] 
\includegraphics[width=0.75\textwidth]{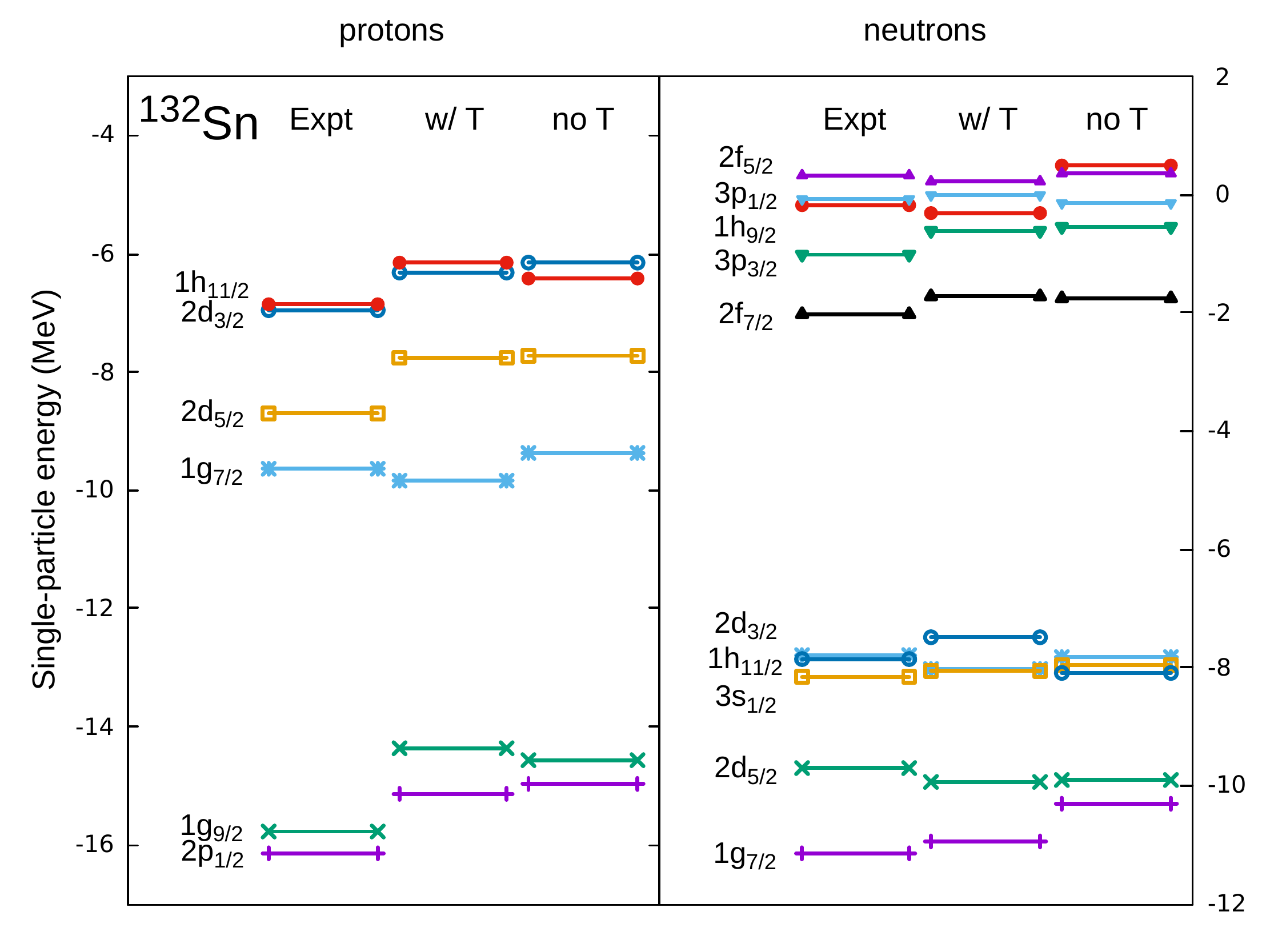}
\caption{Same as in Figure~\ref{Ca40_sps} but for $^{132}$Sn.}
\label{Sn132_sps}
\end{figure}

We now look at tensor effects in the single-particle states of some doubly magic nuclei where data is available. Figures~\ref{Ca40_sps} to~\ref{Sn132_sps} show the single-particle energies calculated from \qmc{III} for the cases with and without the tensor component for $^{40,48}$Ca, $^{56,78}$Ni, and $^{100,132}$Sn, respectively. 

For $^{40}$Ca, which is spin-saturated, tensor effects are expected to be small at sphericity. This is seen in Figure~\ref{Ca40_sps}, where single-particle levels for both proton and neutron states of $^{40}$Ca are not changed with the addition of the tensor component. For both cases, the $2s_{1/2}$ level is pushed up in the QMC results, thereby creating a gap at $Z,N=14$. The gaps at $Z,N=28$ are also pronounced so that the $1f_{7/2}$ state is pushed down, thereby decreasing the proton and neutron shell gaps at $Z,N=20$. The low shell gap has been encountered not just in the current version but was also present in \qmc{II}, as well as in the Skyrme-type forces SV-min and UNEDF1~\cite{frib}.

For the proton states of $^{48}$Ca in Figure~\ref{Ca48_sps}, there is a very slight change in the levels but the effect of tensor terms starts to be distinguishable for its neutron states. The same problem is encounted as in $^{40}$Ca, where there are low proton and neutron shell gaps which are slightly more visible with the tensor component in the neutron states. For the case with tensor, the gap at $N=40$ between the $1f_{5/2}$ and $1g_{9/2}$ shells is larger, so that the $2p$ states are pushed down. The same is true for neutron states of $^{56}$Ni in Figure~\ref{Ni56_sps} where the gap between the $2p_{1/2}$ and $1g_{9/2}$ states, creating the $N=40$ closure, is larger when the tensor component is added.

For $^{78}$Ni in Figure~\ref{Ni78_sps}, the proton and neutron shell gaps from QMC start to pick up so they are now closer to experiment. The $N=40$ gap is still larger if the tensor component is present but for the shell gaps at $Z=28$ between $1f$ states and $N=50$ between $1g_{9/2}$ and $2d_{5/2}$ states, the addition of the tensor component improves the values in comparison with the case where it is omitted.

For $^{100}$Sn in Figure~\ref{Sn100_sps}, there is not much change in the proton states with the addition of the tensor component and the shell gap at $Z=50$ is consistent with experiment. For neutron states, the  gap between states $1g_{9/2}$ and $2d_{5/2}$ is squeezed for both cases, with and without the tensor component, so that it is slightly smaller than that found experimentally.

For $^{132}$Sn proton states in Fig.~\ref{Sn132_sps}, the $Z=50$ gap is again squeezed compared to that of experiment and a small gap forms between $1g_{7/2}$ and $2d_{5/s}$ states. For the neutron case, however, the results reproduce the experimental values quite well. One notable advantage of having the tensor component is that it corrects the order of the $1h_{11/2}$ and $2d_{3/2}$ proton states and the $1h_{9/2}$ and $2f_{5/2}$ neutron states for the $^{132}$Sn isotope.

Clearly the addition of tensor terms in \qmc{III}-T did not lead to any overall improvement in the SO splittings and shell gaps  of the doubly-magic isotopes. As noted in Section~\ref{theory:tensor}, there are strong cancellations in the tensor terms in both the central and spin-dependent parts of the \qmc{III}-T EDF, so that we were not expecting much change from its inclusion. This is also the case for the single-particle spectra, since we did not include SO splittings in the fit data. In most nuclear models, the tensor terms are fitted with additional parameters to control its effect and are tuned to a number of SO splittings of doubly-magic nuclei. The parameters are usually written as $\alpha$ and $\beta$ for the like-particle and proton-neutron tensor component, respectively~\cite{Sagawa2014}. If we rewrite the tensor expressions in Eq.~(\ref{eq:centralJ}) and (\ref{eq:SOJ}), we can identify the corresponding equations for the like-particle and p-n tensor component from \qmc{III}-T as
\begin{eqnarray}
\alpha &=& \frac{G_\sigma(1-dv_0)^2}{2m_\sigma^2} - \frac{G_\omega}{2m_\omega^2}-\frac{G_\rho}{8m_\rho^2} + \frac{1}{8 M^2}\left(-G_\sigma + G_\omega + G_\rho/4 \right) \, ,\\
\beta &=& -\frac{G_\rho}{8m_\rho^2} + \frac{G_\rho}{32M^2} \, .
\end{eqnarray}

Upon comparison of the $\alpha$ and $\beta$ values in Table~\ref{tab:tensor}, we can see that the tensor strength that we have found within the \qmc{III}-T model is relatively small and of opposite signs in comparison with those cases where the tensor parameters in Skyrme forces have been fitted, such as SLy4T~\cite{Zalewski2008}, SLy5+T~\cite{Bender2009} and UNEDF2~\cite{Kort2013}. The SV-min variant which includes a tensor component, SV-tls~\cite{Klup2009}, also has the opposite sign compared to the other Skyrme forces. We highlight, however, that the tensor contribution we find within \qmc{III}-T is fully expressed in terms of the QMC parameters and has not been separately tuned to fit data.
\begin{table}[th!]
\begin{ruledtabular}
	\caption{Comparison of tensor strengths (in MeV$\cdot$fm$^5$) for like-particle $\alpha$ and proton-neutron $\beta$ tensor component from \qmc{III}-T and Skyrme forces SV-tls~\cite{Klup2009}, SLy4T~\cite{Zalewski2008}, SLy5+T~\cite{Bender2009} and UNEDF2~\cite{Kort2013}.}
	\centering
	\begin{tabular} { c c c c c c}
				&\qmc{III}-T	&SV-tls	&SLy4T	&SLy5+T	&UNEDF2\\\hline
               $\alpha$	& 55.2 		&71.1 	&-105	& -89.8 	& -120.3\\
               $\beta$	& -6.3 		&-35.1	&15		& 51.9 	& 11.5 
	\end{tabular}
\label{tab:tensor}
\end{ruledtabular}
\end{table}

Overall, the \qmc{III}-T model tends to emphasise the major shell closures so that some of the states end up being squeezed or pushed higher compared to experiment. This may be because all of the nuclei included in the fit are semi-magic isotopes and isotones, so that closures are mostly emphasised in the fit. It is noteworthy, however, that even if there are no single-particle data included in the fitting procedure, the \qmc{III}-T results do replicate the experimental data quite well. As noted in Ref.~\cite{Sagawa2014}, most nuclear models are either good in terms of their predictions for ground-state bulk properties or single-particle states but not usually for both; the success in one is at most times, at the expense of the other.

\subsubsection{Deformations}
\begin{figure}[th!] 
\includegraphics[width=1.0\textwidth]{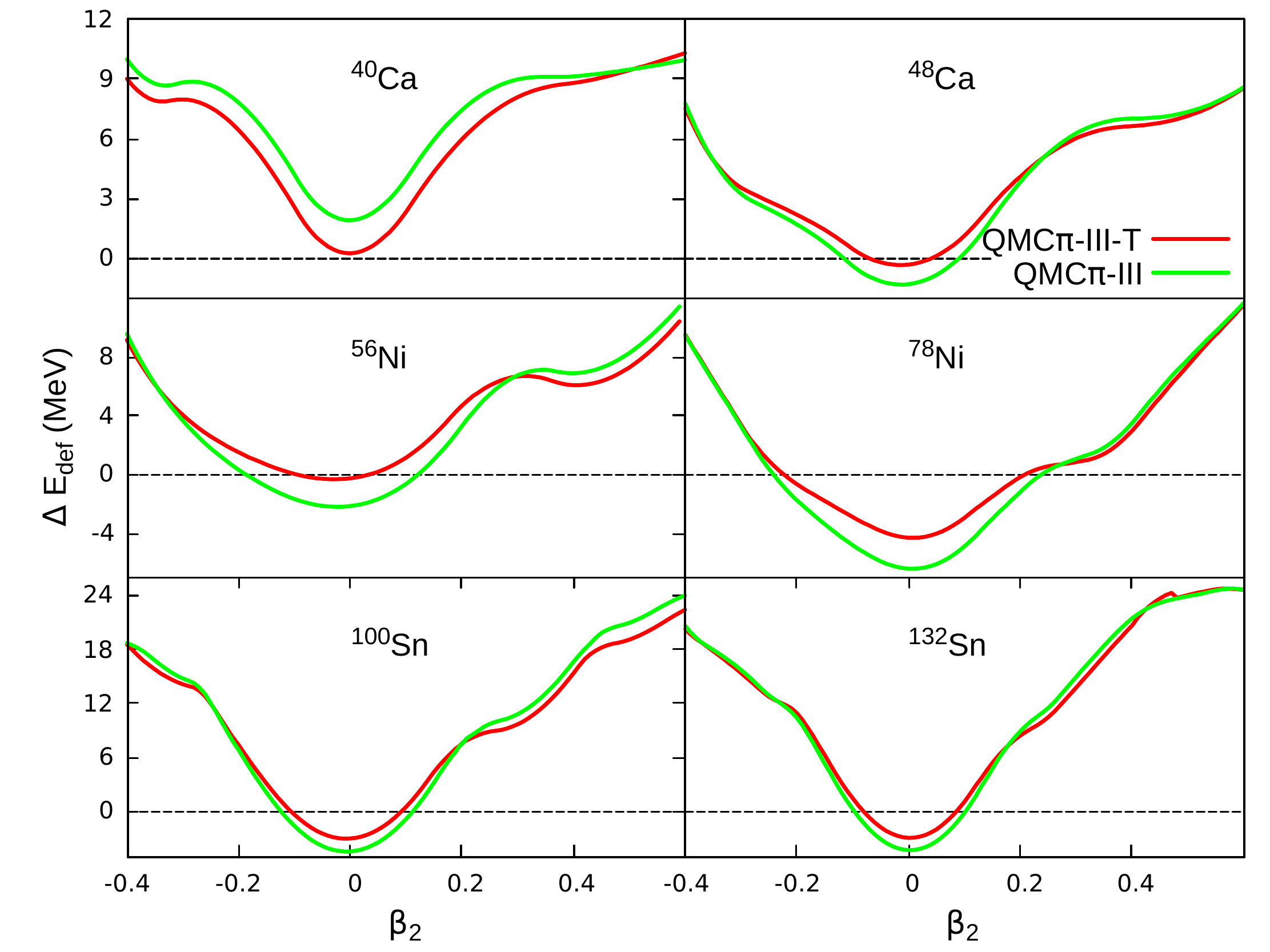}
\caption{Difference in deformation energies from experimental energies (shown in dashed lines) for chosen doubly-magic nuclei, plotted against the deformation parameter $\beta_2$ for the cases with and without a tensor contribution. The plot legend is located in the top right panel.}
\label{deform_doubly}
\end{figure}

The effect of the tensor component on nuclear deformation has been studied in Skyrme EDFs for magic and semi-magic nuclei~\cite{Bender2009,Shi2017} and using the Gogny interaction for medium-mass isotopic chains up to zirconium~\cite{Bernard2016}, where results from various parametrisations were compared. In this section we discuss the contribution of the tensor component to nuclear shapes using the \qmc{III}-T functional. Figure~\ref{deform_doubly} shows the effect of the tensor component on the deformation energy, $E_{\text{def}}$, curves of doubly-magic nuclei. In the figure, $E_{\text{def}}$ values are normalised to the experimental binding energies (shown in dashed lines) and plotted against deformation parameter $\beta_2$.

As mentioned in the previous section, for $^{40}$Ca which is spin-saturated, the tensor effect is expected to have little effect at sphericity. While this is true for the single-particle spectra of $^{40}$Ca, it is not true for the total energy, as seen in Figure~\ref{deform_doubly}. There appears a constant difference between the energy curves with and without the tensor component, which tends to decrease only at large deformation. We emphasise, however, that just as shown in Figure~\ref{fig:BEresid}, the addition of the tensor term improved the $BE$ value for the doubly-magic and symmetric $^{40}$Ca isotope and thus its minimum in Figure~\ref{deform_doubly} is closer to that of the experimental value.

For systems that are not spin-saturated, like  $^{56,78}$Ni, and $^{100,132}$Sn, tensor effects are expected to dominate only around sphericity. This can be seen in Figure~\ref{deform_doubly} for these nuclei, where curves with and without the tensor term tend to behave in a similar way as the deformation increases; the difference occurs mostly around sphericity and decreases towards $^{132}$Sn. Again, we emphasise that the minima for these doubly-magic nuclei are closer to experimental data when the tensor component is added, as also shown in the $BE$ residuals in Figure~\ref{fig:BEresid}.

Deformation plots and tensor effects were also studied along the zirconium chain, where shape transitions are expected as $N$ increases from the spherical $^{90}$Zr. Table~\ref{b2Zr} lists the ground-state $\beta_2$ deformations for the Zr chain for the two cases of \qmc{III}, values from FRDM and Skyrme forces SV-min and UNEDF1, along with available data for some Zr isotopes. The tensor component did not significantly change the ground-state deformation along the Zr chain for both cases in \qmc{III}, as the $\beta_2$ values are almost the same. From the spherical $^{90}$Zr isotope, deformation slightly increases to the oblate side up to $^{96}$Zr, while isotopes switch shape to being highly prolate starting from $^{98}$Zr up to $^{112}$Zr. The QMC results are consistent with FRDM for heavy Zr isotopes as well as with available data, which suggests that $^{100}$Zr to $^{106}$Zr have highly prolate shapes. For some Skyrme parametrizations in Refs.~\cite{Bender2009, Shi2017} the main impact of the tensor component is the disappearance of the deformed minimum of $^{100}$Zr. Note that this is not the case for QMC, where the deformed minimum did not change much along the Zr isotopic chain, even with the inclusion of the tensor component.
\begin{table}[th!]
	\centering
	\caption{Deformation parameter $\beta_2$ for Zr ($Z=40$) isotopes. FRDM results are taken from \cite{Moller2016} while SV-min and UNEDF1 are taken from \cite{frib}. Experimental data for $N\le50\le62$ are taken from \cite{Raman2001} and for $N=64,66$ data are from \cite{Browne2015}.}
\begin{ruledtabular}
	\begin{tabular} {c c|c c c c c c}
N&A&\qmc{III}-T&\qmc{III}&FRDM&SV-min&UNEDF1&Expt\\
\hline
50	&	90	&	-0.04&	-0.03&	0.00	&	0.00	&	0.00	&	0.09	\\
52	&	92	&	-0.14&	-0.13&	0.00	&	0.00	&	0.00	&	0.10	\\
54	&	94	&	-0.17&	-0.16&	-0.16	&	0.00	&	0.00	&	0.09	\\
56	&	96	&	-0.19&	-0.18&	0.24	&	0.00	&	0.00	&	0.08	\\
58	&	98	&	0.51	&	0.50	&	0.34	&	0.00	&	0.00	&	no data	\\
60	&	100	&	0.45	&	0.44	&	0.36	&	-0.18	&	-0.18	&	0.36	\\
62	&	102	&	0.46	&	0.46	&	0.38	&	0.37	&	0.38	&	0.43	\\
64	&	104	&	0.45	&	0.45	&	0.38	&	0.37	&	0.38	&	0.39(1)	\\
66	&	106	&	0.44	&	0.44	&	0.37	&	0.37	&	-0.20	&	0.36(1)	\\
68	&	108	&	0.42	&	0.42	&	0.36	&	-0.19	&	-0.20	&	no data	\\
70	&	110	&	0.43	&	0.41	&	0.36	&	0.00	&	0.00	&	no data	\\
72	&	112	&	0.48	&	0.47	&	0.36	&	0.00	&	0.00	&	no data	\\
	\end{tabular}
	\label{b2Zr}
\end{ruledtabular}
\end{table}

\begin{figure}[th!] 
\includegraphics[width=1.0\textwidth]{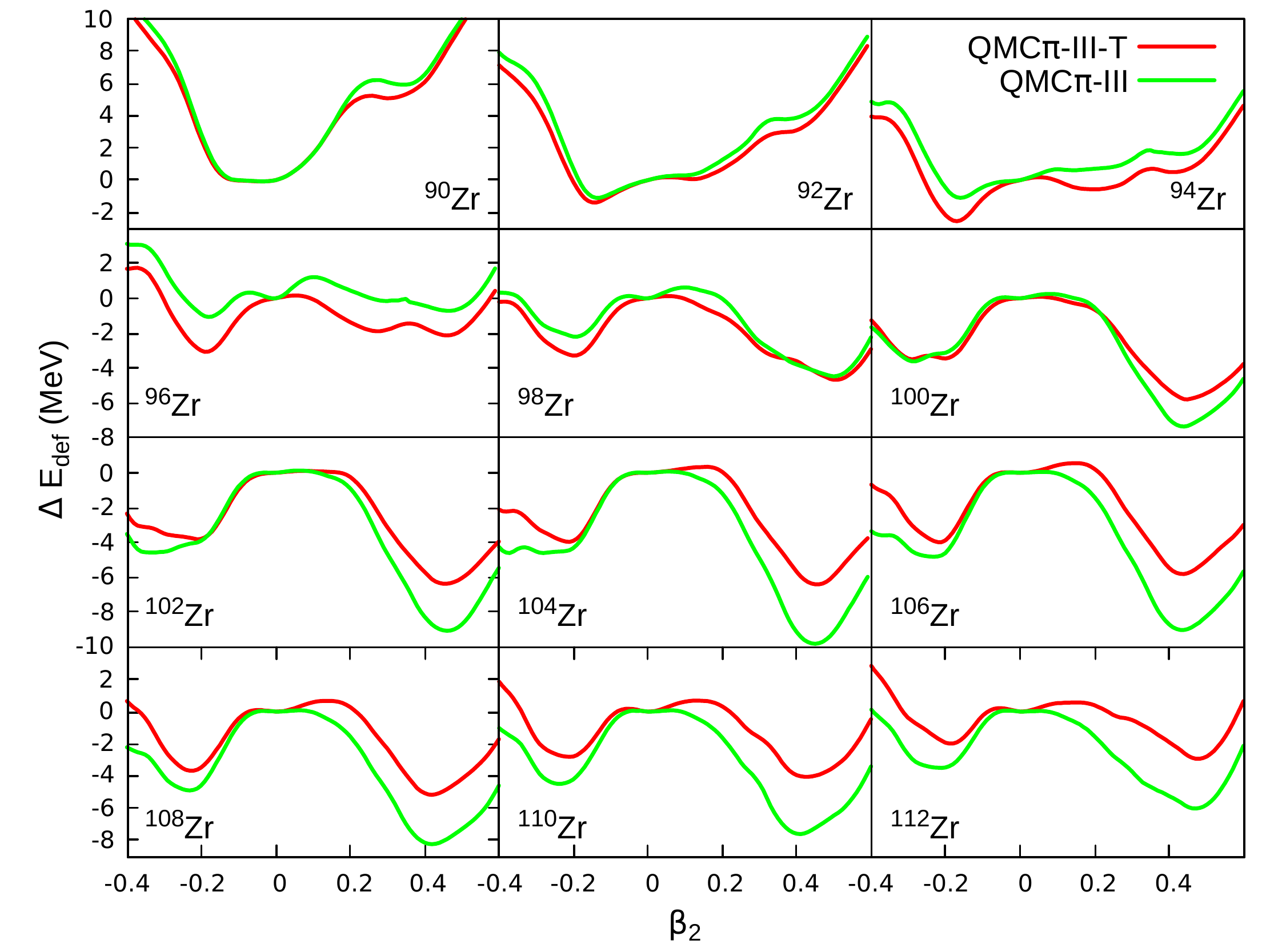}
\caption{Difference in deformation energy from the value at sphericity plotted against deformation parameter $\beta_2$ for the Zr chain for the cases with and without the tensor term in \qmc{III}. The plot legend is located in the top right panel.}
\label{Zr_deform}
\end{figure}

Figure~\ref{Zr_deform} shows the deformation energy plots for the Zr chain from $A=90$ to 112, comparing results from \qmc{III} with and without the tensor component. For the spherical $^{90}$Zr, the effect of the tensor component with deformation is only appreciable at around $\beta_2\approx 0.3$, where there is a flatter shoulder compared to the case without tensor. Although somewhat flat, the minimum for $^{92}$Zr starts to shift to the oblate side with a $\beta_2$ value of around -0.14. From $A=94$ to 96, the effect of the tensor component is to yield deeper minima, so that both nuclei appear to be oblate. However, there seems to be a triple shape coexistence for the case without tensor in $^{96}$Zr, where minima at $\beta_2\approx-0.2,0$, and 0.4 almost have the same energy values. Starting from $^{98}$Zr, the first minimum shifts to the prolate side, although the second minimum, which is oblate, still has a deformed energy close to that of the first prolate minimum. For both cases, with and without the tensor term, the shape evolution across the 
$\beta_2$ values for $^{98}$Zr is almost the same, contrary to those of the previous two Zr isotopes where tensor effects are slightly pronounced. The prolate minima continues to exist from $A=100$ to 112 but this time the case without tensor develops deeper minima in contrast to those of the lighter Zr isotopes. Further, the prolate minimum starts to shift up from $^{108}$Zr and as $A$ increases so that, at $^{112}$Zr, the deformed minimum in the prolate and oblate side balances out when the tensor component is present.

\subsection{Pairing functionals and QMC}
In the earlier versions of QMC for finite nuclei, we employed nuclear pairing throughout the nuclear volume using a $\delta$-function force (DF). In this subsection, we compare results from \qmc{III}-T with a density-dependent pairing functional (DDDI) to that of \qmc{III}-T with DF pairing, as discussed in Section \ref{sub:QMCpair}. Also added for comparison are results from the previous version \qmc{II}, where DF pairing was also employed.

\begin{figure}[th!] 
\includegraphics[width=1.0\textwidth]{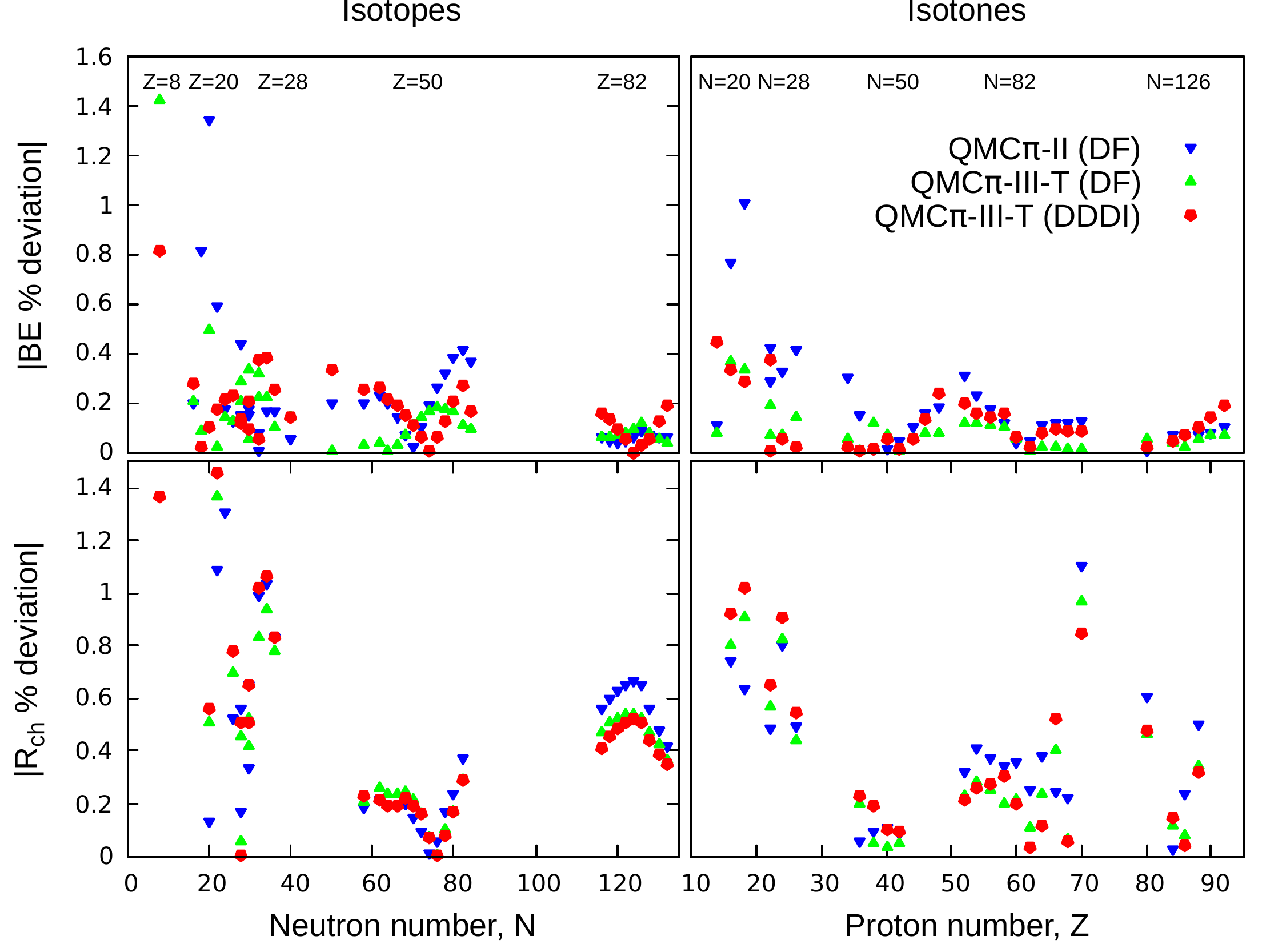}
\caption{Percentage deviation from experiment for binding energies $BE$ and \textit{rms} charge radii $R_{ch}$ for the 70 nuclei included in the fit for various choices of pairing interaction, as explained in the text. The plot legend is located in the top right panel.}
\label{fig:pair}
\end{figure}

Figure~\ref{fig:pair} shows a comparison of fit results from \qmc{II} and \qmc{III}-T with different pairing functionals. It should be emphasised that the \qmc{III}-T (DF) functional, just like that in the \qmc{II} case, contains two extra pairing strength parameters, as in many other mean-field models. In general, the two pairing functionals within the \qmc{III}-T model tend to have similar fit results for binding energies and charge radii. The only noticeable difference is found in the energies of neutron-deficient Sn isotopes, where DF performs slightly better. Compared to the results from \qmc{II}, we can see an overall improvement with the current version, especially in the energies of the Ca and Sn isotopes and in the isotonic chains. We also see improvements for charge radii, especially for the Pb isotopes.

\begin{figure}[th!] 
\centering
\includegraphics[width=1.0\textwidth]{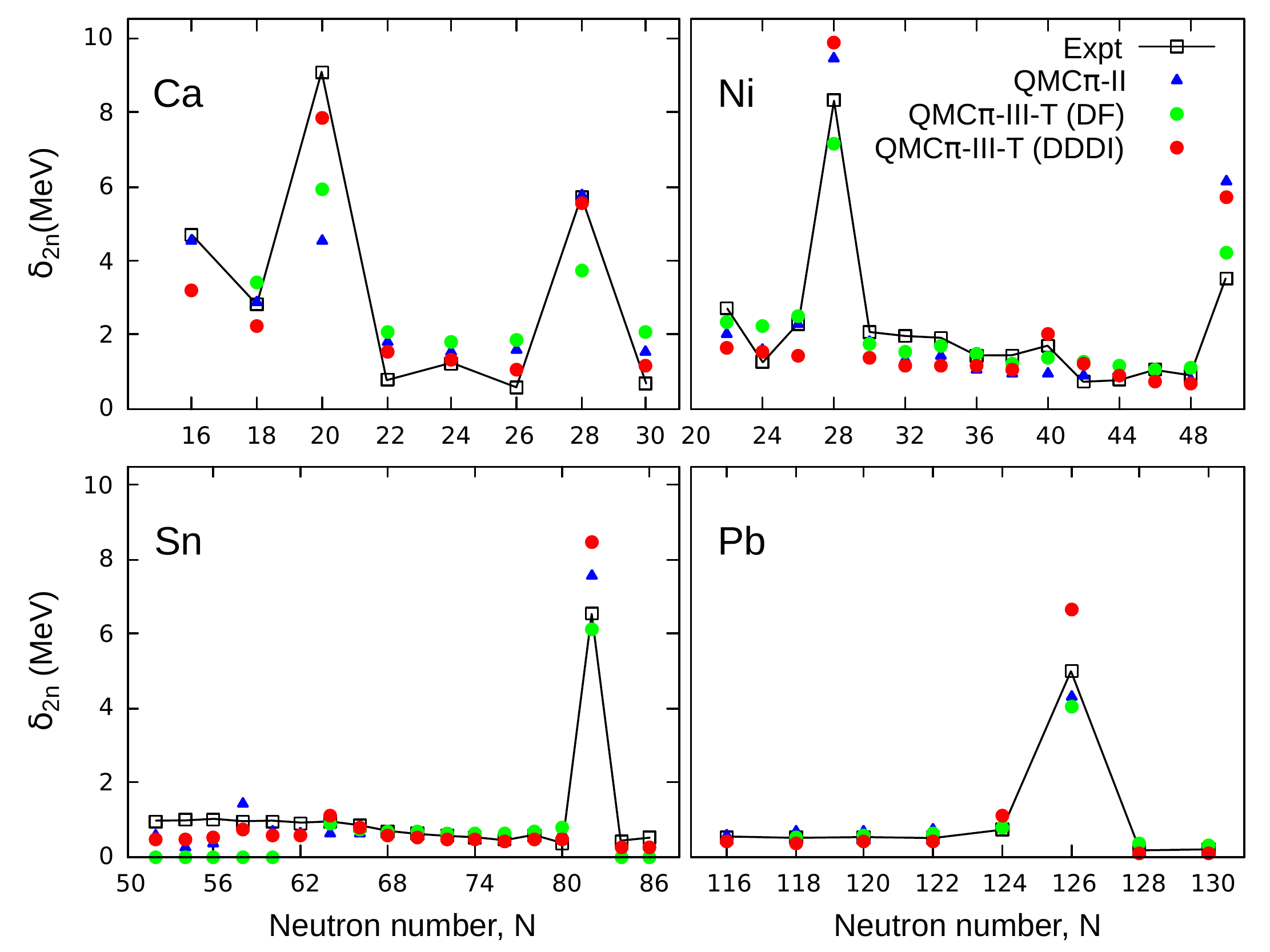}
\caption{Two-neutron shell gap for Ca, Ni, Sn, and Pb isotopes computed from QMC model with different pairing functionals. Experimental data are taken from masses in~\cite{Wang2017} and data points are connected by lines to emphasise the peaks at shell closures. The plot legend is located in the top right panel.}
\label{fig:shelln_isotopes}
\end{figure}

We highlight a significant feature of the density-dependent QMC-derived pairing which relates to the predictions for shell closures. Figure~\ref{fig:shelln_isotopes} shows the two-neutron shell gaps, $\delta_{2n}$, for Ca, Ni, Sn, and Pb isotopes computed from the QMC model with different pairing functionals. Peaks in shell gaps are signatures of shell or subshell closures as can be seen in the magic neutron numbers. For Ca isotopes, \qmc{III}-T with DDDI pairing gives a very good description of the shell closures at $N=20$ and 28, while the model tends to overestimate the values for Ni, Sn and Pb at $N=28, 50, 82$ and 126. Nevertheless, the $\delta_{2n}$ peaks are very appreciable in these magic numbers and it is significant that only \qmc{III}-T (DDDI) is able to replicate the relatively small peak corresponding to the closure at $N=40$ in the Ni chain. 

\begin{figure}[th!] 
\centering
\includegraphics[width=1.0\textwidth]{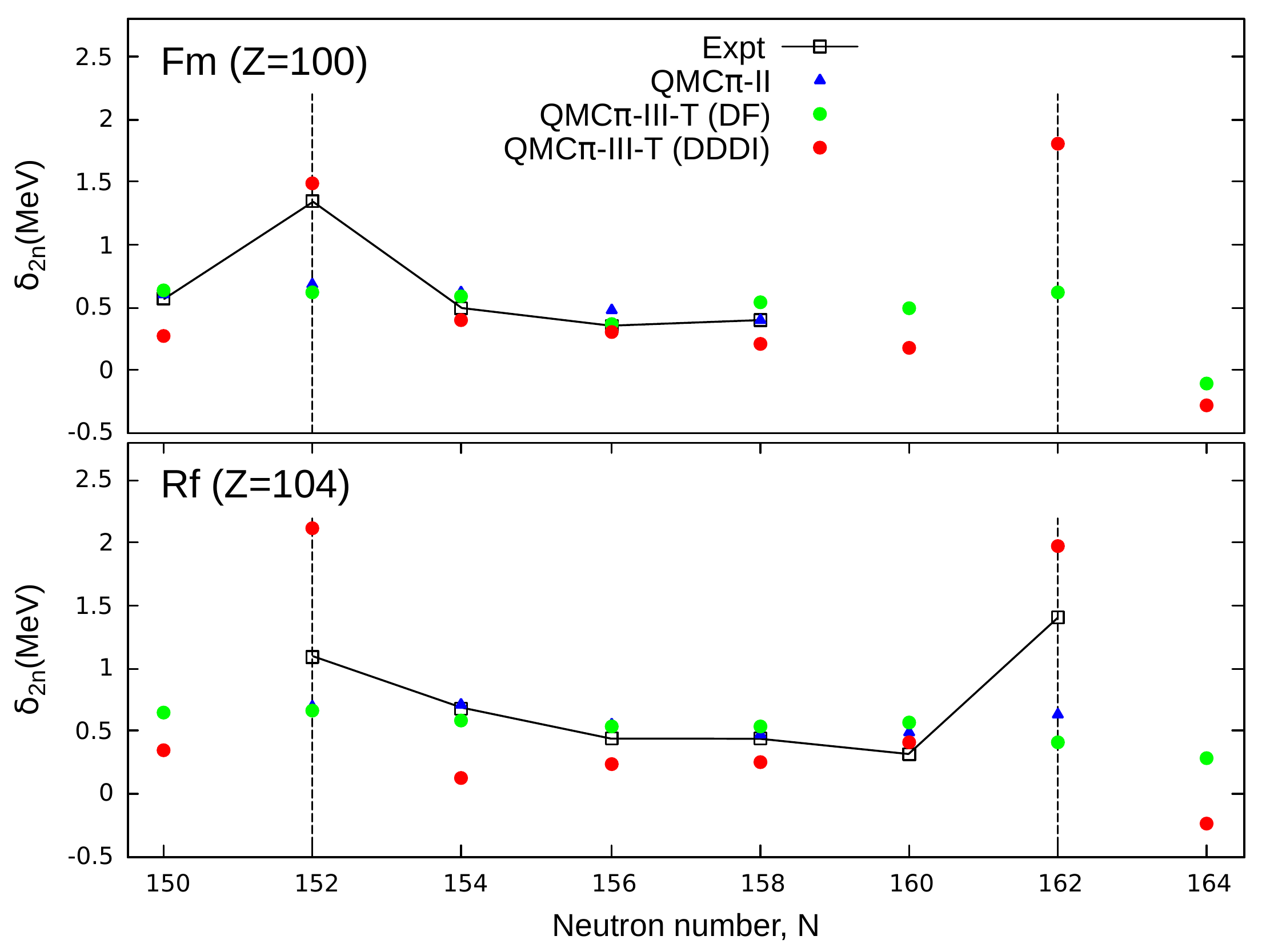}
\caption{Same as in Fig.~\ref{fig:shelln_isotopes} but for Fm ($Z=100$) and Rf ($Z=104$) isotopes.}
\label{fig:shelln_SHE}
\end{figure}

A very interesting result for shell gaps is seen in the superheavy region where DF pairing fails to reproduce the peaks at $N=152$ and $N=162$ which have been seen in experiment. Figure~\ref{fig:shelln_SHE} shows the $\delta_{2n}$ values for the Fm ($Z=100$) and Rf ($Z=104$) isotopic chains. Though somewhat overestimated, the $\delta_{2n}$ peaks are well reproduced by \qmc{III}-T with DDDI pairing and are absent for \qmc{II} and \qmc{III}-T with DF pairing. This suggests that pairing should be taken to be density-dependent, especially in SHE, if one is to replicate the observed subshell closures and provide better predictions for possible closures higher up the nuclear chart, where experimental data is not yet available. Subshell closures and other predictions in the superheavy region will be discussed in a separate writeup.
%

\section{Concluding remarks}
\label{conclude}
The latest \qmc{III}-T EDF has been improved by the addition of the tensor component which naturally arises from the model, a pairing functional that has been derived within the QMC framework and a full expression for the \s Hamiltonian contribution. With these developments, the overall level of agreement with the observables for finite nuclei improved significantly, particularly for the binding energies and radii, compared to the previous \qmc{II} version. These results are of a similar quality to those found in other modern energy density functionals, despite the reduction in the total number of parameters in the current model. Moreover, the resulting nuclear matter parameters changed very little from the values found in the previous version, \qmc{II}, which lie well within the acceptable ranges.

The effect of adding the tensor terms in QMC mostly improved the total $BE$ of the isotopes and isotones included in the fit, while they had little effect on the single-particle spectra. While their contribution was expected to improve the spin-orbit splittings and shell gaps, this was not the case for \qmc{III}-T, as the effect of the tensor terms was rather small. It is emphasised, however, that we did not fit any new parameters for the inclusion of the tensor component and that we did not include single-particle data in the fit in this current version; the strength of tensor component was solely determined by the combination of QMC parameters which were fitted solely to $BE$ and $R_{ch}$. Furthermore, while the tensor component does not change the sphericity of doubly-magic isotopes and the deformed shapes of neutron-rich Zr isotopes, its effect can be seen in the energy curves plotted against deformation parameter $\beta_2$, by shifting the minima up or down, thereby creating flatter or deeper minima. 

Another improvement in the current version appears in the pairing functional, where the pairing parameters are now expressed in terms of the meson-nucleon couplings. With the resulting density-dependent pairing, the shell closures for medium to heavy and most importantly the subshell closures in superheavies are now replicated well, in comparison with the results from having volume pairing that was used in the older QMC versions. More calculations and discussions in the superheavy region using the latest \qmc{III}-T will be presented in future work.

\section*{Acknowledgements}

J. R. S. and P. A. M. G. acknowledge with pleasure the support and hospitality of the CSSM at the University of Adelaide during visits in the course of this project. This work was supported by the University of Adelaide and by the Australian Research Council through Discovery Projects DP150103101 and DP180100497.

\bibliography{Paper2} 
\end{document}